\documentclass[sn-mathphys]{sn-jnl}

\jyear{2021}%

\theoremstyle{thmstyleone}%
\theoremstyle{thmstyletwo}%
\usepackage{multirow}

\theoremstyle{thmstylethree}%

\usepackage{verbatim} 
\usepackage{tabularx} 
\usepackage{lineno}
\begin{document}

\title[A novel nuclear recoil calibration for liquid helium detectors]{A novel nuclear recoil calibration for liquid helium detectors}

\author*[1]{\fnm{Fengbo} \sur{Gu}}\email{FengboGu@outlook.com}

\author[1]{\fnm{Jiangfeng} \sur{Zhou}}\email{853051644@qq.com}

\author*[1,2]{\fnm{Junhui} \sur{Liao}}\email{junhui\_liao@brown.edu}

\author[3]{\fnm{Yuanning} \sur{Gao}}\email{yuanning.gao@pku.edu.cn}

\author[1]{\fnm{Zhuo} \sur{Liang}}\email{liangzhuo\_w@163.com}

\author[1]{\fnm{Meiyuenan} \sur{Ma}}\email{1321351097@qq.com}

\author[1]{\fnm{Zhaohua} \sur{Peng}} \email{pzh44@sina.com}

\author[4]{\fnm{Lifeng } \sur{Zhang}}\email{zlf20042008@126.com}

\author[4]{\fnm{Lei} \sur{Zhang}}\email{zlamp@163.com}

\author[1]{\fnm{Jian} \sur{Zheng}}\email{13522656935@139.com}

\affil[1]{\orgdiv{Division of Nuclear Physics}, \orgname{China Institute of Atomic Energy}, \orgaddress{\street{Sanqiang Rd. 1}, \city{Fangshan district}, \postcode{102413}, \state{Beijing}, \country{China}}}

\affil*[2]{\orgdiv{Department of Physics}, \orgname{Brown University}, \orgaddress{\street{Hope St. 182}, \city{Providence}, \postcode{02912}, \state{Rhode Island}, \country{USA}}}

\affil[3]{\orgdiv{School of Physics}, \orgname{Peking University}, \orgaddress{\street{ChengFu Rd. 209}, \city{Haidian district}, \postcode{100084}, \state{Beijing}, \country{China}}}


\affil[4]{\orgdiv{Division of Nuclear Synthesis Technology}, \orgname{China Institute of Atomic Energy}, \orgaddress{\street{Sanqiang Rd. 1}, \city{Fangshan district}, \postcode{102413}, \state{Beijing}, \country{China}}}

\abstract{According to many dark matter models, a potential signal registered in a detector would feature a single-scattering nuclear recoil (NR). So, it is crucial to calibrate the detector's response to NR events. The conventional calibrations implement $\sim$ keV to  MeV neutrons, which can be produced by an accelerator, a neutron generator, or a radioactive source. Although the calibrating methods have been widely employed, they could be improved in several ways: (a) the incident neutron energy should be more monoenergetic, (b) the calibrating NR energy should line up with the  region of interest (ROI) of the experiment, and (c) the intensity of the beam should be appropriate. In the paper, we introduce a novel NR calibration method for liquid helium detectors, in which a helium beam ($\alpha$ particles) will be implemented to calibrate the detectors. The helium beam can (i) be tuned precisely to have a jitter of $\lesssim $ 4\% (the $\alpha$  beam's kinetic energy is equivalent to the recoil energy in the conventional calibrations with fast neutrons); (ii) have an energy between $\sim$ 100 eV and tens of keV; and (iii) provide a tunable flux from nA to 100 $\mu$A, which presents convenience in beam pipe configuration to obtain a $\sim$ 100 Hz events rate so that the events pileup would be ignorable.}


\maketitle

\section{Nuclear recoil calibration for Dark Matter detectors}\label{sec1}
\subsection{Introduction to nuclear recoil calibration}\label{sec1sub1}

Weakly Interacting Massive Particles (WIMPs) are a class of dark matter candidate that result in a signal consisting of (a) single-scattering~\footnote{Since the recoil energy is only $\sim$ keV to MeV scale so that it would be an elastic scattering.} and (b) nuclear recoil (NR). It is a single-scattering instead of multi-scattering because the cross-section between WIMPs and the detector's nucleons~\footnote{A nucleon refers to a neutron or a proton. Essentially, the interaction between a WIMPs particle and a nucleon is the WIMPs coupling with the quarks of the nucleon.} is at the scale of the weak interaction or smaller. Therefore, if WIMPs interact with matter, they would hit a meters-size detector only once. The NR feature comes from the assumption that WIMPs would scatter coherently with the detector's nucleus (neutrons and protons combined) and can not generate electron recoil (ER) events directly because they are not supposed to participate in an electromagnetic interaction.

Ideally, one should calibrate a dark matter detector with WIMPs beams. Due to the lack of such a beam, physicists implement fast neutrons (with $\sim$ keV to MeV kinetic energy) to do calibration. Regarding a fundamental interaction, a neutron interacts with the detector strongly, while a WIMPs particle has a yet-to-known interaction with the detector. So, the fundamental interaction of the two calibrating methods is different. However, they are equivalent in the sense that the responses of the detector are the same: the detector's nuclei will recoil and generate observable signals accordingly.

The NR calibration essentially characterizes the ionizing signal of a detector when the nuclear recoil registers energy inside. The detector's readout system can measure the signal of ionization. The nuclear recoil energy is often extrapolated from neutrons, i.e., the neutrons' kinetic energy difference before and after scattering the target detector. The neutron's initial energy can be obtained from the beam facility~\cite{Burke74, Gerbier90, Messous95, Simon03, Jagemann05,  Izraelevitch17, Bonhomme22} and the scattered neutron's energy is often measured by a time of flight (TOF) detector. In addition to accelerators, deuterium-deuterium (D-D) neutron generators~\cite{Verbus17, Huang22} and radioactive sources~\cite{LZTDR17} can also produce fast neutrons.

A neutron beam cannot be bent and boosted like other charged particle beams. So, the energy of neutron beams can not be tuned. 
In many cases, the uncertainties of the neutron energy will eventually show up as the error bars of the calibrating results. 
As mentioned in reference~\cite{Verbus17}, for the 2.45 MeV neutron generated by the DD108 neutron generator (made by Adelphi Technology, Inc.), the intrinsic width ($\sigma/\mu$) of the outgoing neutron energy distribution is $\lesssim$ 5\%. Reference~\cite{Lang18} tested an NSD/Gradel-Fusion neutron generator and found the neutron energy was roughly 2.2 to 2.7 MeV. Often, a neutron detector and the affiliated TOF system have a more significant uncertainty than the energy of the outgoing neutrons produced by a D-D generator. Effectively, the D-D generator can be considered a monoenergetic neutron source. However, to calibrate the liquid helium (LHe) detector for dark matter searches, $\sim$ keV to tens of keV neutrons are needed. So, if one still wants to use a D-D generator for the calibration, the MeV neutrons must be moderated to 2 to 3 orders lower energy. One way to obtain the keV neutrons with a D-D generator is to put one or more reflectors at specific angles. As simulated in reference~\cite{Liao2018}, assuming the incident neutrons energy is fixed to be 2.45 MeV, the energy of the neutrons scattered from the reflector ($\sim$ 20 cm away) would have a $\sim$ 2\% uncertainty (1 $\sigma$). Moreover, the to-be-calibrated detector's intrinsic energy resolution will contribute to the recoil energy's uncertainty and eventually to the NR calibrating results. Another concern is the beam's intensity. According to the simulation~\cite{Liao2018}, 1 billion (D-D generator produced) 2.45 MeV neutrons would only have $\sim$ 150 reflected $\sim$ 300 keV neutrons from the reflector. The flux of the scattered (and desired) neutrons is 7 orders smaller than the generator initially produced. With two reflectors, $\sim$ 30 keV neutrons can be obtained, but the flux would be 1/10$^{14}$ of the generator. The Adelphi DD109 neutron generator can generate neutrons up to 10$^9$/s. So it would need 10$^5$ seconds or $\sim$ 1 day to have one $\sim$ 30 keV neutron. Taking 1000 events needs to run the machine for three years. So, the ``D-D generator + reflectors'' configuration is not ideal for us to calibrate the LHe detectors at $\sim$ keV scale.

Alternatively, one can implement an ion beam to calibrate a detector, as the COMIMAC facility demonstrated~\cite{Santos08, Muraz16, newsG2022}. With the facility, the beam can be tuned to have an energy uncertainty as low as $\lesssim$ 4 \%~\cite{newsG2022}, which is the total uncertainty of the recoil energy. In the conventional calibrations, the uncertainty mainly contributed from the energy of the incident neutrons and the scattered neutrons. It could be a factor of a few (and more) greater than 4 \%.

The paper will be organized as follows. In section~\ref{sec2}, we will introduce the COMIMAC facility; we further address the scheme of calibrating an LHe detector with the COMIMAC facility in section~\ref{sec3}; next, we explain why the same scheme might not be suitable for other liquid noble gas detectors such as liquid argon (LAr) and liquid xenon (LXe) in section~\ref{sec4}; in section~\ref{sec5}, after discussing the pros and cons of the calibrating method, we conclude that the calibrating method we introduced here could be complementary to other ones already available in the community.

\section{The COMIMAC facility}\label{sec2}
The COMIMAC facility is a table-size ion source developed by the MIcro-tpc MAtrix of Chamber (MIMAC) team at Laboratoire de Physique Subatomique et de Cosmologie (LPSC) in Grenoble, France. The facility has an ion source called COmpact MIcro-wave and Coaxial (COMIC) to produce ions and electrons from a gas, a Wien filter to  select desired particles based on their charge/mass ratios, two pieces of deflectors to adjust the beam vertically, a Faraday cup measures the beam current, and an interface wall having a $\sim$ 1 to 13 $\mu$m  hole in the center through which the beam transmit to the detector to be calibrated downstream. For details, please refer to~\cite{Muraz16, newsG2022}. 
The facility was originally designed to calibrate gas detectors. We are planning to implement the device to calibrate liquid helium detectors~\cite{discussionWithCOMIMACGuys}, as will be discussed in details below. 

The facility can produce two beams: the ion of the filled gas and the electron beam. This paper mainly focuses on the ion beam for NR calibration, though the electron beam can also serve for ER calibration. The kinetic energy of the ion beam can be expressed as E$_{\text{kin}}$ = q $\cdot$ U$_{field}$, where q and U$_{field}$ are the number of elementary charges of the ion and the electric field being applied to the system, respectively. So, in principle, the uncertainty of the beam's kinetic energy would only depend on the electric power supply, which often has an ignorable jitter. Reference~\cite{newsG2022} compared the signal between the COMIMAC-generated electrons and an X-ray generator-produced keV photoelectrons; a conservative uncertainty of 4\% on the beam's kinetic energy was estimated. The uncertainty is at least a factor of a few smaller than the recoil energy obtained by the conventional methods with fast neutrons. The intensity of the beam flux can also be tuned between a few nA to 100 $\mu$A, which gives convenience for users to reach a $\sim$ 100 Hz event rate~\cite{newsG2022}. 
In addition, the beam energy can be tuned from $\sim$ 100 eV to 50 (10) keV if the beam emits horizontally (vertically). The 100 eV$_{nr}$ - 50 keV$_{nr}$\footnote{Here, keV$_{nr}$ represents the nuclear recoil energy in keV. A related  notation is keV$_{ee}$, which is electron equivalent recoil energy. A $\sim$ keV to MeV neutron or a WIMPs particle hits a detector will generate $\sim$ keV nuclear recoil energy, keV$_{nr}$, part of which will convert to electron equivalent recoil energy to ionize the detector, keV$_{ee}$. The ratio of the two energy, keV$_{ee}$ / keV$_{nr}$, is often called an ionization efficiency or a quenching factor.} aligns with the ROI of the liquid helium detectors aiming for dark matter searches.

\section{Calibrating an LHe detector with COMIMAC}\label{sec3}
Researchers at Brown University have proposed to hunt for Solar neutrinos with liquid helium (HERON)~\cite{HERON1} and dark matter~\cite{HERON88, HERON96} in the 1990s or so. Recently, some new ideas have been developed in detector design and signal readout~\cite{Maris17, Hertel19, Biekert22}. LHe detectors have also been implemented for a neutron electric dipole moment (nEDM) experiment~\cite{Huffman2000, Ito2012, Ito2016, Ito16, Phan20}. Although the concept of an LHe time projection chamber (TPC) was discussed almost ten years ago~\cite{GuoMckinsey13}, it has yet to be built (for dark matter hunting). ALETHEIA, which stands for A Liquid hElium Time projection cHambEr In dArk matter, would be the first collaboration aiming to hunt for dark matter and beyond with LHe-filled TPCs~\cite{ALETHEIA-EPJP-2023}.

We have launched a series of R\&D programs on LHe detectors at China Institute of Atomic Energy (CIAE) since 2020 and have made significant progress ever since. So far, we have demonstrated that a single-phase LHe TPC is technologically viable~\cite{ALETHEIA-EPJP-2023, TPBCoutingJINSTNov22, TPBCoutingALETHEIA22, MaFBKSIPMPDE23}, and the verification of a dual-phase LHe detector is currently underway~\cite{ALETHEIA2023dualPhaseTPC}. In addition, we have confirmed that the cryogenic type of FBK SiPMs~\cite{FBKCompanyLink} can work at $\sim$ 7 K~\cite{ALETHEIAPDE2023}, the photo-detection efficiency  (PDE) reaches 40\% at the temperature when a 9 V overvoltage (OV) has been applied, and the typical after-pulse and cross-talk rates are both around 10\% when the OV is lower than $\sim$ 10 V. Thanks to the single-digit number of ER and NR backgrounds on a 100-kg scale LHe TPC, the detector can touch down the neutrino floor/fog~\cite{Billard14, OHare21} with 3-year data taking. In addition to the conventional ``ER excess only'' channel and the ``NR excess only'' channel, an LHe TPC can search for the possible dark matter signals in the form of ``ER excess + NR excess''. According to our knowledge, ALETHEIA is the first project to be dedicated to the search of any form of dark matter signals (ER excess only, NR excess only, or ER + NR excess combined)~\cite{ALETHEIA2023dualPhaseTPC}. 

Helium is significantly lighter than other noble gases such as argon and xenon, so an LHe TPC has the advantage of hunting for low-mass WIMPs, provided it can be successfully built. On the NR channel, the ALETHEIA's ROI corresponds to $\sim$ 0.2 to 10 keV$_{nr}$, which perfectly lines up with the COMIMAC facility's beam energy, 0.1 eV$_{nr}$ - 10 keV$_{nr}$ (when the beam is vertical). As to ER calibration, in addition to implementing the COMIMAC facility's electron beam, immersing a $^{63}$Ni source into the liquid helium to characterize the detector could be another option, as demonstrated in reference~\cite{Sethumadhavan06}. The $^{63}$Ni is an electron source with an endpoint energy of 66 keV and an average energy of 17 keV~\cite{Ni63SourceIAEA}. 


\subsection{Calibrating a liquid helium detector with COMIMAC}\label{sec3sub1}
As mentioned in references~\cite{Santos08, Muraz16}, the LPSC team has successfully demonstrated that the COMIMAC facility and its predecessor can calibrate helium gas's ionization efficiency (or quenching factor) with minimal uncertainties. References~\cite{newsG2022} measured the ionization efficiency of protons in methane. The paper also analyzed the uncertainty of the beam energy to be $<$ 4\%.

There are a few unique engineering challenges to calibrating an LHe detector with the COMIMAC facility. (a) The beam pipe must seal well with several cryogenic shielding layers to prevent vacuum and heat leakage. (b) LHe must be in the superfluid status to have the lowest possible vacuum above so the beam can hit directly on LHe instead of encountering the helium gas above the liquid. (c) Since superfluid helium atoms will creep up along the inner surfaces of the container and eventually go into the beam pipe to stop the beam, a dedicated mechanical structure must be implemented to avoid that happening. (d) In our application, the beam must travel a 30 cm length pipe before hitting the target, so the beam's direction, energy, intensity, and secondary electrons will differ from the LPSC team~\cite{Santos08, Muraz16, newsG2022}. However, detailed analysis, calculation, and simulation have convinced us that we can calibrate LHe detectors with the beam generated by the COMIMAC facility.


\begin{figure}[!t]	 
	\centering
    \includegraphics[width=1.0\textwidth]{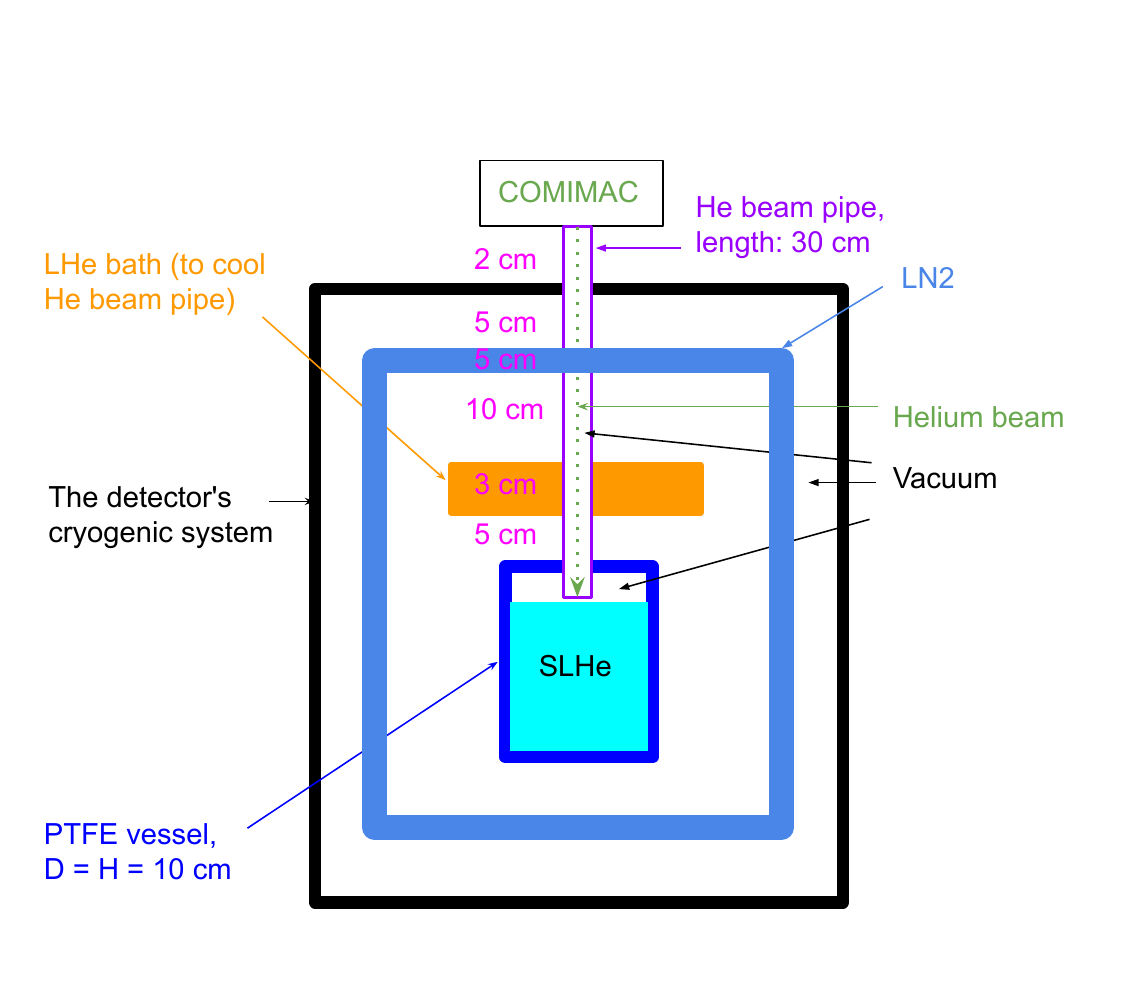}
	\caption{The schematic drawing shows the implementation of the COMIMAC facility to do an NR calibration on a liquid helium detector: the beam goes through the 30 cm pipe to calibrate the SLHe detector down there. The distance or thickness of every part along the pipe is marked. Please refer to the main text for more info.}\label{LHeCalibraionCOMIMAC} 
\end{figure}

As schematically shown in Fig.~\ref{LHeCalibraionCOMIMAC}, the COMIMAC beam goes through the 30 cm copper pipe all the way down to the superfluid liquid helium (SLHe) detector. We chose copper as the pipe material mainly because it has an ignorable ratio of secondary electrons; please refer to section~\ref{sec3sub3} for more info. The 2 K LHe bath aims to help the beam pipe's cold end hold at 0.5 K. The top part of the SLHe detector is a $\sim$ 1 cm thick vacuum space. Since the beam is always in a vacuum environment before hitting the SLHe target, the energy loss of the beam would be minimal. A more detailed analysis is in the following.

\begin{figure}[!t]	 
	\centering
    \includegraphics[width=1.0\textwidth]{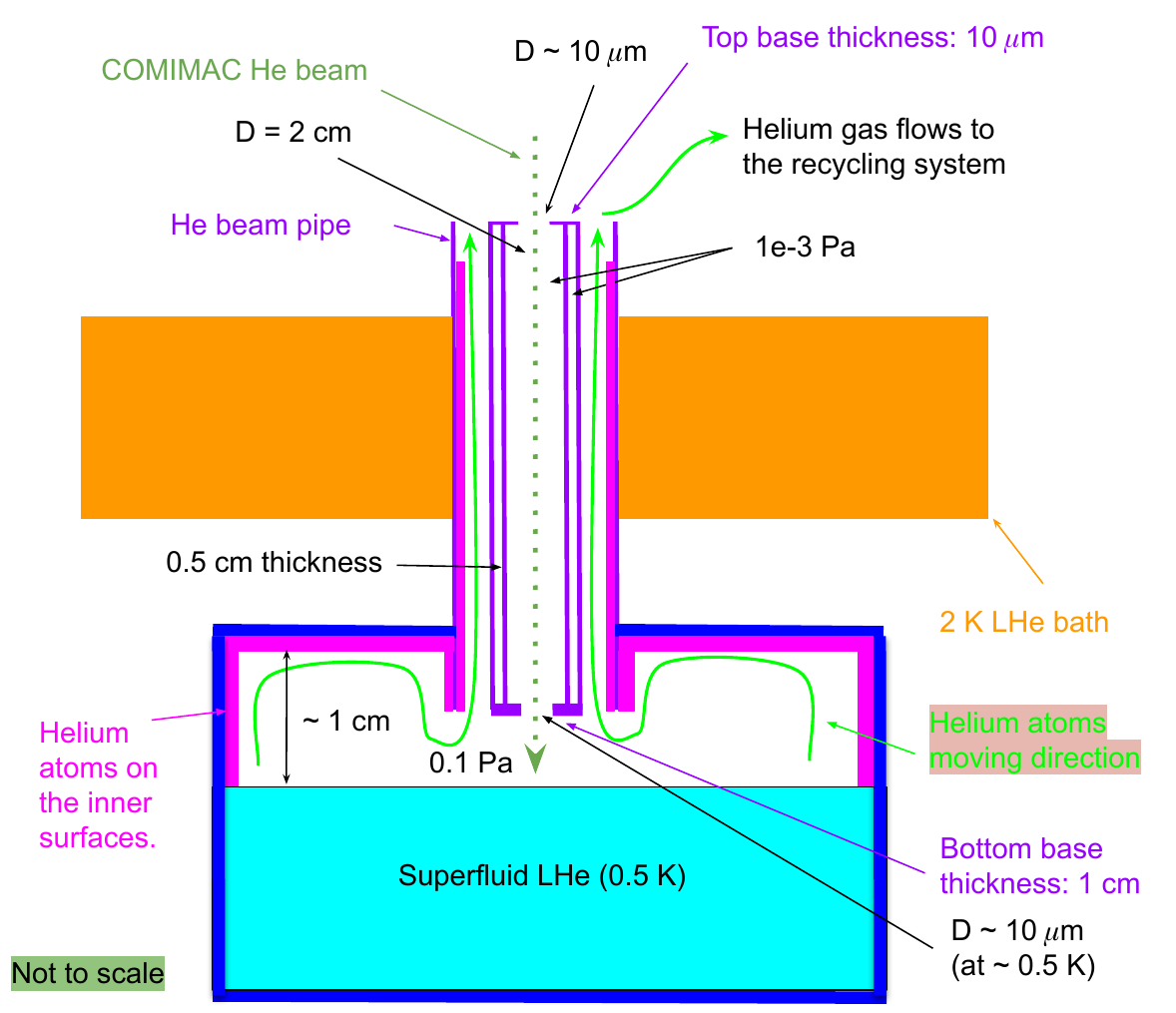}
	\caption{The detailed view of Fig~\ref{LHeCalibraionCOMIMAC}. The 30 cm length copper pipe has a 10 $\mu$m diameter on the top base to match up with the COMIMAC facility. The pipe has two layers, each with a 1e-3 Pa vacuum inside. The inner layer is 2 cm in diameter to let the beam pass through; the outer layer is $\sim$ 0.5 cm in thickness to prevent heat leakage. The bottom base of the pipe is 1 cm thick, and the diameter of the hole is 10 $\mu$m at 0.5 K (The diameter would be $\sim$ 10.03 $\mu$m at room temperature). Since helium atoms can climb up along the inner walls at the temperature and evaporate later, a special structure is designed to split the incoming beam and the vaporized helium gas into two separate tunnels so the beam can go into the SLHe area without losing energy. For more info, please refer to the main text.}\label{LHeCalibraionCOMIMACDetailedView} 
\end{figure}

As we can see in Fig.~\ref{LHeCalibraionCOMIMACDetailedView}, the COMIMAC helium beam goes through the 30 cm pipe and the $\sim$ 1.0 cm height, 0.1 Pa vacuum space, then impinges the SLHe. The pipe has a 10 $\mu$m diameter aperture on the top base to match the COMIMAC facility. The pipe has two layers, each with a 1e-3 Pa vacuum inside. The inner layer is 2 cm in diameter to let the beam pass through; the outer layer is $\sim$ 0.5 cm thick to prevent heat leakage. The bottom base of the pipe is 1 cm thick, and the diameter of the hole is 10 $\mu$m at 0.5 K (The diameter would be $\sim$ 10.03 $\mu$m at room temperature). The pipe's bottom base is effectively a collimator for the beam. please refer to section~\ref{sec3sub3} for details. 

The temperature determines the vacuum of the 1.0 cm height space. In principle, the lower the temperature of LHe, the lower the vacuum of the space~\cite{Donnelly98}. So, in an actual cryogenic setup as Fig.~\ref{LHeCalibraionCOMIMACDetailedView}, if the temperature of the SLHe is below 0.5 K, the vacuum of the 1.0 cm height space would be less than 0.1 Pa. The chance of an incident particle being scattered in the area would be smaller than when the SLHe is 0.5 K.

Helium gas has a boiling point of 4.2 K and a lambda point of 2.17 K~\cite{Donnelly98}. Below  2.17 K, LHe changes to a superfluid phase to become SLHe, which has many different properties to general LHe. The most relevant one might be that superfluid helium can creep up along the inner walls of a chamber against the force of gravity. In the setup shown in Fig.~\ref{LHeCalibraionCOMIMAC}, without a special mechanical design, the SLHe atoms will climb up the pipe's inner walls then vaporize at a certain place where the temperature is above 4.2 K.  If this happens, the incident helium beam will be ``absorbed'' entirely by the vaporized helium gas.
Accordingly, we designed a beam pipe with two nested pipes, as shown in Fig.~\ref{LHeCalibraionCOMIMACDetailedView}. The inner one has a diameter of 2 cm, which would allow the helium beam to go through. The out one is intentionally designed to let SLHe atoms creep and eventually flow to the helium recycling system (a molecular pump is connected). There is a 0.5 cm thickness vacuum layer between the two nested pipes to reduce heat leakage. In the figure, the pink strips represent the creeping SLHe atoms, which will move upwards along the pipe walls, as the green lines indicate.

\subsection{The beam's kinematic energy loss}\label{sec3sub2}

We employed three independent methods to understand the beam's energy loss before hitting the target SLHe, i.e., the energy deposited in the 30 cm length, 1e-3 Pa air beam pipe, and the $\sim$ 1.0 cm thickness, 0.1 Pa helium gas on the top of the chamber (above the SLHe): (a) A calculation based on National Institute of Standards and Technology (NIST) data~\cite{heliumStoppingPower}, (b) A simulation with SRIM~\cite{SrimWebsite}, and (c) A Geant4~\cite{Geant4Website} simulation. The consistent results were obtained with the three methods, as shown in the following. 

We calculated the energy loss with the stopping power of helium ions in helium according to the NIST data~\cite{heliumStoppingPower}, as shown in the NIST columns of table~\ref{tabSumED}. In the calculation, the incident particle is helium, and we assume the stopping power of the helium ions does not change during the journey. The assumption is reasonable because the energy loss is only at $\sim$ eV level, which is ignorable for keV particles. 

The calibration with the SRIM is similar to the method with the NIST data. We first simulated the dE/dx of the helium ions with SRIM on three energies: 2 keV, 10 keV, and 50 keV; we then calculated the energy deposition by assuming the dE/dx is constant. The calculated results are shown in the SRIM columns in table~\ref{tabSumED}. The NIST and SRIM results are almost the same.

In the Geant simulation, we assumed (a) the beam was monoenergetic and (b) the beam was fine enough so that it did not touch the pipe's surfaces during the journey (except the 1e-3 Pa air and the 0.1 Pa helium gas). The simulation with other beam configurations will be discussed in section~\ref{sec3sub3}. The Geant version is 4.11.1.0. The physics list is ``FTFP\_BERT''. The ``RegisterPhysics'' for the electromagnetic model is not the default one of the FTFP\_BERT, ``G4EmStandardPhysics'', but the ``G4EmLowEPPhysics'', which employs the low energy Livermore model for the bremsstrahlung and ionization process of electrons and uses the Urban model for the multiple Coulomb scattering~\cite{Anh21, Geant4Website}. For 1 MeV and below energy, the EM process has been replaced with a nuclear stopping model. We simulated the deposited energy in the 30 cm, 1e-3 Pa air, and the 0.1 Pa helium gas with different thicknesses, 0.5 cm, 1.0 cm, and 2.0 cm. The results are shown in the ``G4'' columns in table~\ref{tabSumED}. Generally speaking, the Geant4 simulated results are consistent with the ones calculated with NIST and SRIM. However, the 2 keV beam discrepancies are relatively more significant than 10 keV and 50 keV beams. Given the typical value of (deposited energy) / (beam energy) is $\sim$ 0.1\%, the beam's energy loss in the whole journey would be ignorable provided it passes through the 30 cm pipe and the 0.1 Pa helium gas ``smoothly'' before hitting the SLHe. More details will be introduced in section~\ref{sec3sub3}.

\begin{table}[]
\caption{The G4 columns show the mean deposited energy in the 30 cm length, 1e-3 Pa air beam pipe, the 0.1 Pa helium gas space of different thicknesses, and the SLHe. The NIST columns  represent the calculated energy deposition based on the NIST stopping power data. The SRIM columns are the simulated energy deposition with SRIM.}\label{tabSumED}
   \begin{tabular}{c c c c c c c c c c}
\toprule
        \multicolumn{1}{c}{\multirow{2}{*}{\textbf{Medium}}}  & \multicolumn{3}{c}{\textbf{2 keV $\alpha$}}      & \multicolumn{3}{c}{\textbf{10 keV $\alpha$}} 	&\multicolumn{3}{c}{\textbf{50 keV $\alpha$}} \\ \cline{2-10}
        \multicolumn{1}{c}{length}   		& \textbf{G4}    &\textbf{NIST}      &\textbf{SRIM}   &\textbf{G4} & \textbf{NIST}   &\textbf{SRIM} 	 &\textbf{G4} & \textbf{NIST}   &\textbf{SRIM} \\
        \multicolumn{1}{c}{(cm)}                              & (keV) 		&(keV)              	& (keV)		& (keV) 		&(keV)    	&(keV) 		 	& (keV) 	& (keV)                        &(keV)  \\ 
\toprule
        \multicolumn{1}{c}{30.0 }   				&5.0e-5           &                     	&9.6e-5		&1.2e-4            &                &1.4e-4   			&2.8e-4	&          			&2.6e-4 \\ 
        \multicolumn{1}{c}{(1e-3 Pa air)}      	        &             		&                       &			&                      &                & 				&		&           			&\\ 
        \multicolumn{1}{c}{0.5 }    				&1.7e-3           & 2.2e-3             	&2.3e-3		&2.5e-3            &2.6e-3      &2.6e-3   			&5.2e-5  &5.4e-3  			&5.4e-3\\ 
        \multicolumn{1}{c}{(0.1 Pa He$_2$)}      	&             		&                       &			&                      &                & 				&		&           			&\\ 
        \multicolumn{1}{c}{1.0 }     				&3.2e-3           & 4.3e-3             	&4.7e-3		&5.0e-3            &5.2e-3      &5.3e-3  			&1.0e-2  &1.1e-2  			&1.1e-2\\ 
        \multicolumn{1}{c}{(0.1 Pa He$_2$)}      	&             		&                       &			&                      &                &				&		&           			&\\ 
        \multicolumn{1}{c}{2.0 }    				&6.2e-3           & 8.7e-3             	&9.3e-3		&9.9e-3            &1.0e-2      &1.1e-2 			&2.1e-2  &2.2e-2  			&2.2e-2\\ 
        \multicolumn{1}{c}{(0.1 Pa He$_2$)}      	&             		&                       &			&                      &                &				&		&           			&\\ 
        \multicolumn{1}{c}{SLHe}  				&1.6	                &                        &			&9.2                 &                &				&50		&           			&\\ 

\botrule
    \end{tabular}
\end{table}

\begin{figure}[!t]	 
	\centering
    \includegraphics[width=1.0\textwidth]{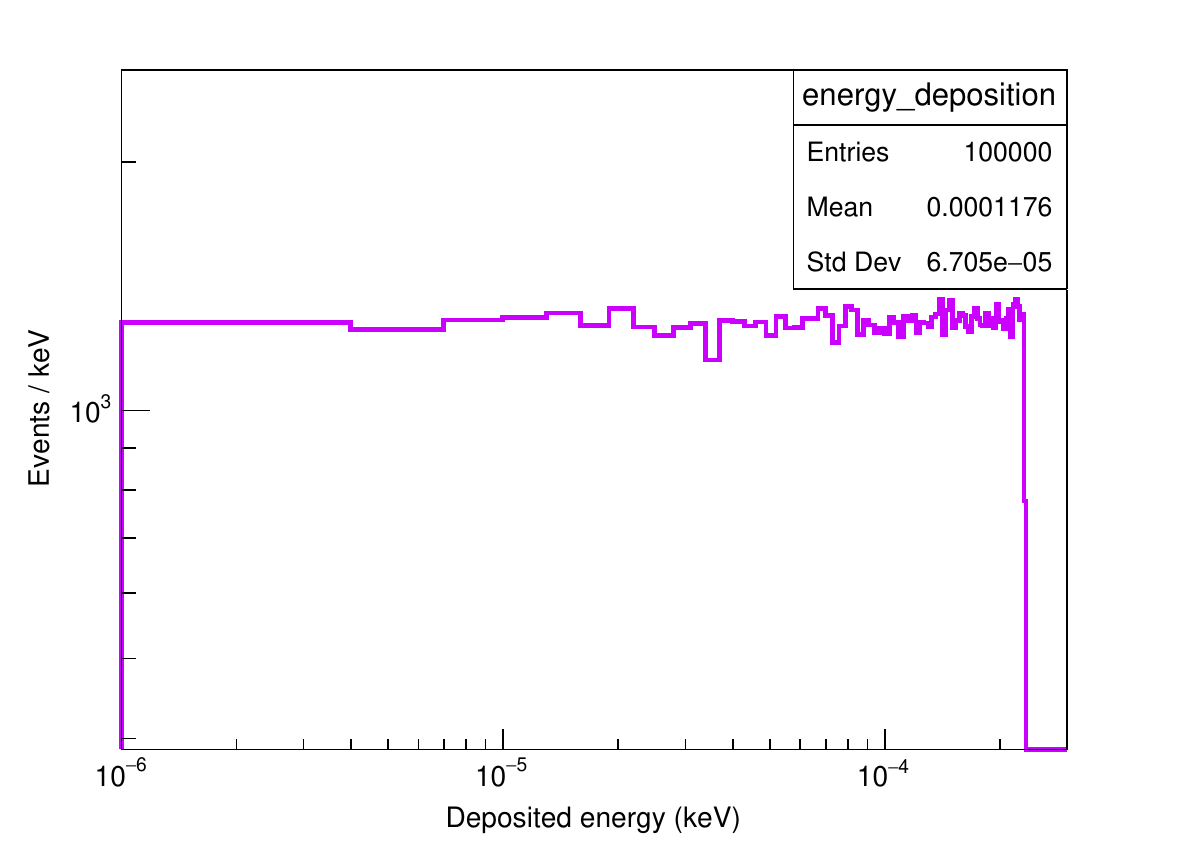}
	\caption{Geant4 simulated energy deposition in the 30 cm length, 1e-3 Pa beam pipe when the 10 keV helium beam passes through. The simulation aims to know the energy deposition in the pipe's vacuum, so the beam was hypothesized to be fine enough not to touch the pipe's surfaces during the journey. A more realistic simulation will be presented in section~\ref{sec3sub3}.}\label{GeantSim50cmPipe} 
\end{figure}

\begin{figure}[!t]	 
	\centering
    \includegraphics[width=1.0\textwidth]{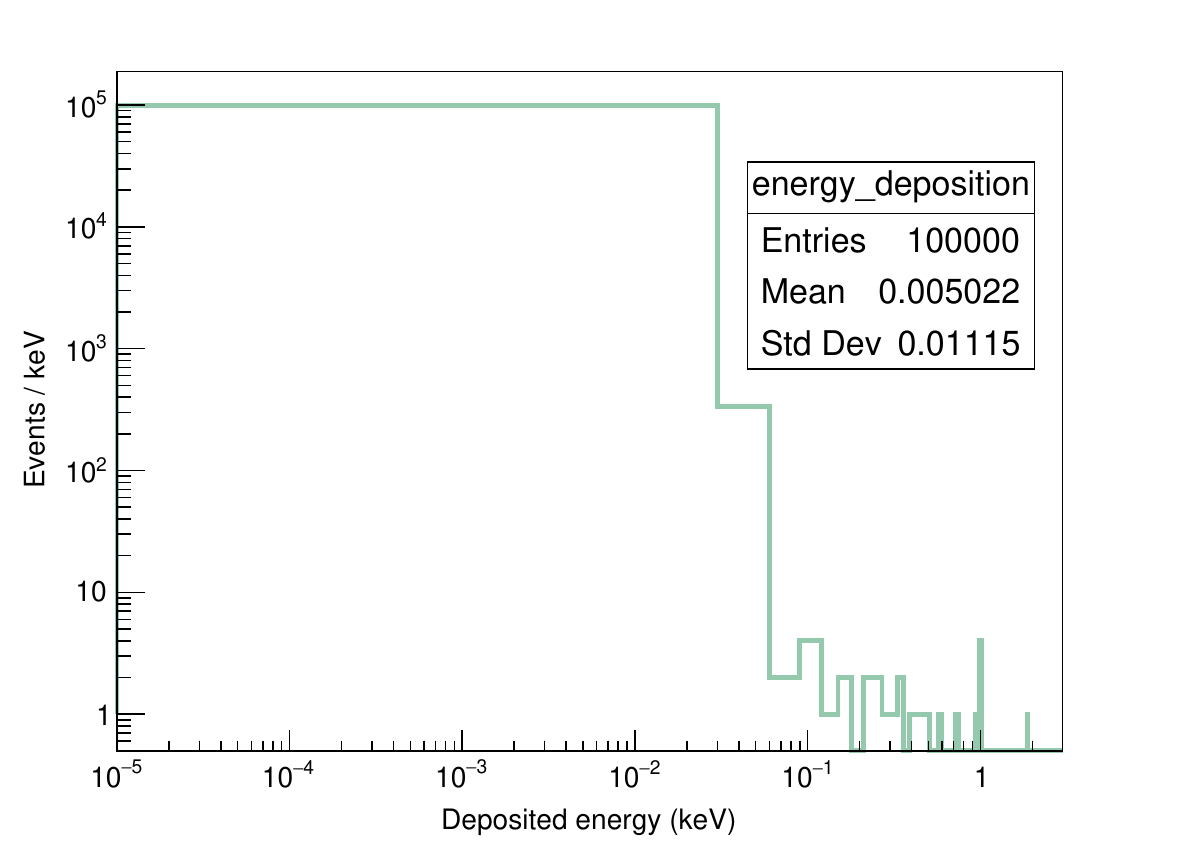}
	\caption{Geant4 simulated energy deposition in the 1 cm thickness, 0.1 Pa helium gas when the 10 keV helium beam passes through. The simulation aims to know the energy deposition in the helium gas layer after going through the 30 cm length vacuum pipe, so the beam was hypothesized to be fine enough not to touch the pipe's surfaces during the journey. A more realistic simulation will be presented in section~\ref{sec3sub3}. }\label{GeantSim1cmHelium} 
\end{figure}

\begin{figure}[!t]	 
	\centering
    \includegraphics[width=1.0\textwidth]{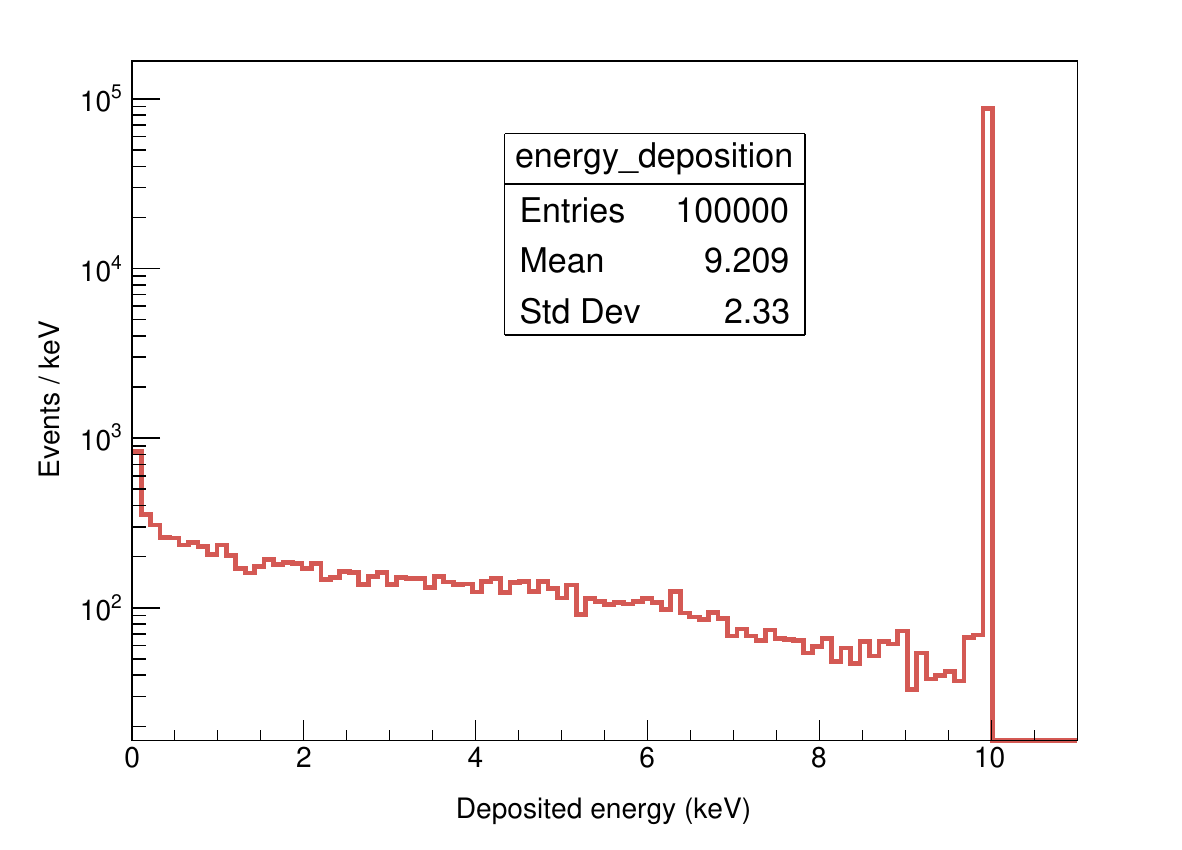}
	\caption{Geant4 simulated energy deposition in the SLHe when the 10 keV helium beam passes through. The simulation aims to know the energy deposition in the SLHe after the beam going through the 30 cm length vacuum pipe and the 1 cm thickness, 0.1 Pa helium gas, so the beam was hypothesized to be fine enough not to touch the pipe's surfaces during the journey. A more realistic simulation will be presented in section~\ref{sec3sub3}.}\label{GeantSimSLHe} 
\end{figure}

As examples, Fig.~\ref{GeantSim50cmPipe}, Fig.~\ref{GeantSim1cmHelium}, and Fig.~\ref{GeantSimSLHe} show the deposited energy in the pipe, the helium gas, and the SLHe with 10 keV beams, respectively. 

\subsection{Studying the deposit energy in the SLHe}\label{sec3sub3}

\begin{figure}[!t]	 
	\centering
    \includegraphics[width=1.0\textwidth]{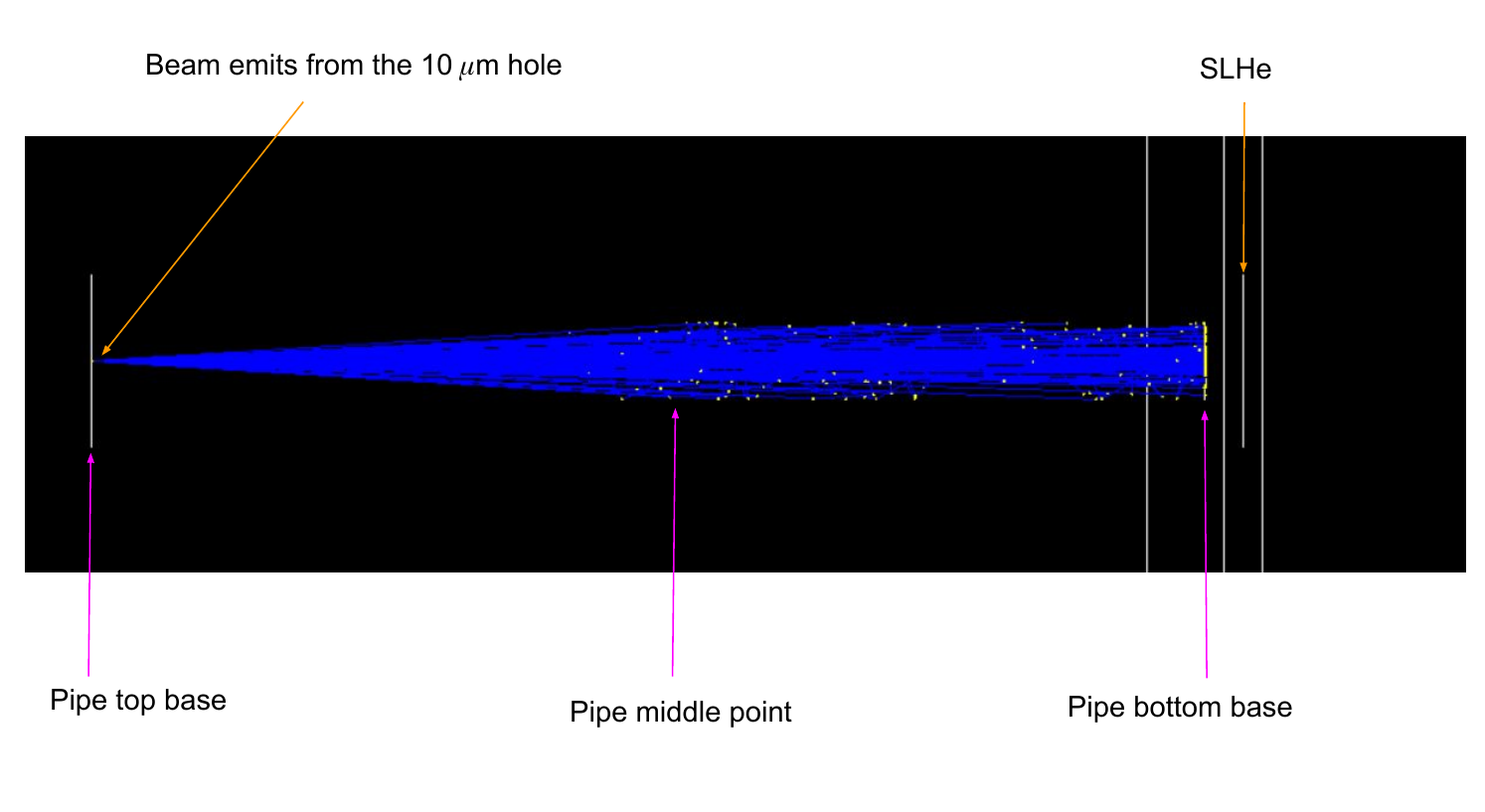}
	\caption{The Geant4 simulation shows the trajectory of 10 keV helium ions emitting from the 10 $\mu$m hole, then hitting the middle point of the 30 cm length beam pipe (on the inner wall).}\label{GeantSimBeamHitPipeMiddle} 
\end{figure}

The beam size from the COMIMAC facility is supposed to be a few mm$^2$, while the apeture on the top base of the pipe is only 10 $\mu$m in diameter. The bottom base is 1 cm thick, having a $\sim$ 10 $\mu$m hole in the center at room temperature (The diameter would shrink to $\sim$ 9.97 $\mu$m at 0.5 K if it is made of copper). The bottom base can be effectively considered as a beam collimator. The precise intensity and shape of the beam impinging on the hole of the top base are unknown. However, given that (a) the pipe is 30 cm in length, and (b) the beam can not be focused after it goes into the pipe, almost all of the beam particles would be absorbed by the inner walls of the pipe, only very few helium ions can go through the 1 cm length, 10 $\mu$m diameter collimator in cryogenic, then hit the liquid helium down there. As an example, we assume the beam impinges the middle point of the pipe (on the inner wall), as shown in Fig.~\ref{GeantSimBeamHitPipeMiddle}. With Geant4, we simulated the deposited energy in the SLHe.

\begin{figure}[!t]	 
	\centering
    \includegraphics[width=1.0\textwidth]{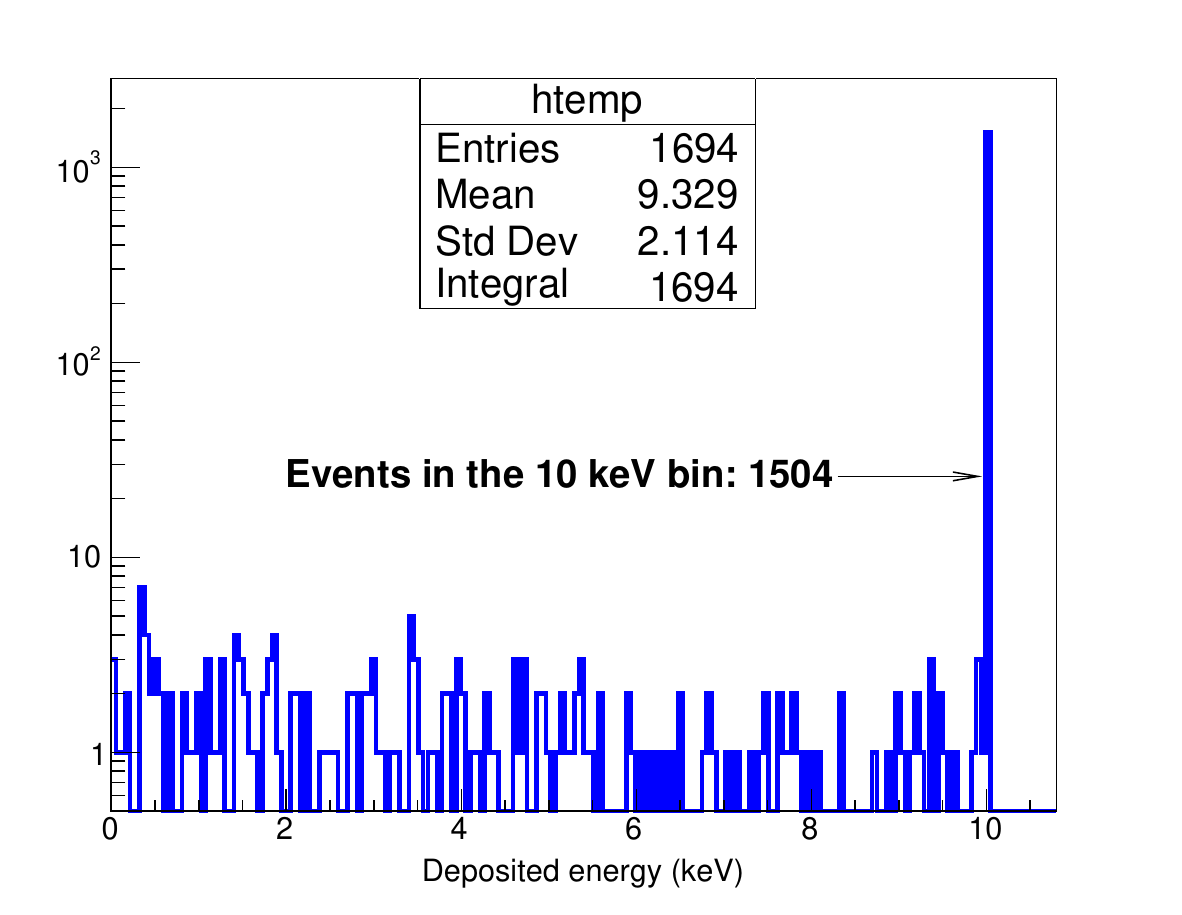}
	\caption{Geant4 simulated energy deposition in the SLHe when 10 M 10 keV helium ions impinged the 10 $\mu$m diameter hole on the top base. Only 1694 particles entered into the SLHe. The number of events in the 10 keV bin is 1504, representing the meaningful calibration data.}\label{GeantSimSLHeKasiRealBeamCollimator} 
\end{figure}

For 10 M incident helium ions, only 1694 events registered in the SLHe; among these events, 1504 of them deposited $\sim$ 10 keV, as Fig.~\ref{GeantSimSLHeKasiRealBeamCollimator} shows. Because $\sim$ keV helium ions are supposed to deposit all of their energy into SLHe, we can reject the 190 (= 1694 - 1504) events with an energy cut. We should also consider the detector's energy resolution in experimental data so the selection would be slightly different. In brief, with such a design, most of the events registered in the SLHe are meaningful; for the lower energy events, we can exclude them with an energy cut. The simulation can not mimic the actual beam entirely but should be representative enough on the R\&D stage. We will keep on optimizing the pipe configuration with simulation and actual data.

As mentioned above, for the 10 M incident helium ions, only 1504 meaningful events registered in the SLHe. The ratio is $\sim$ 10$^{-4}$ ($\approxeq$ 1504/10 M). According to the tests with the COMIMAC facility~\cite{newsG2022}, a $\sim$ 100 Hz events rate can be achieved with a $\sim$ nA beam flux. To achieve the same events rate, we need to increase the beam intensity by a factor of 10$^{4}$ to compensate for the lost particles in our 30 cm pipe. The COMIC ion source is 5 W~\cite{Muraz16}, corresponding to a $\sim$ 0.1 mA for 10 keV He 0$^+$ beams. So, the beam flux can potentially increase by 5 orders (0.1 mA / nA), which is more than enough for us. On the other hand, even if the facility can not increase the beam flux anymore, the event rate would be 0.01 Hz  ($\sim$ 30 events per hour), which is still viable for our calibrations.

Secondary electrons and sputtering are the other background resources for the calibration. The efficiency of sputtering depends on the beam ions and the materials the beam hits. As mentioned in reference~\cite{Buschhaus23}, copper (Cu) has an ignorable secondary electrons events ratio compared to aluminum (Al) and Titanium (Ti), only $\sim$ 0.05 events per ion for 2 keV and 4 keV ions, and the energy of the electrons are only tens of eV. So, we chose Cu as the pipe's material. For 10 M helium ions impinging on the 30 cm copper pipe, the secondary electrons would be 0.5 M, and the number of electrons entering into SLHe would be $\sim$ 50 according to the abovementioned simulation. These 10s eV electron events can be rejected by an energy cut completely.

\subsection{COMOSL simulation on temperatures}\label{sec3sub4}

\begin{figure}[!t]	 
	\centering
    \includegraphics[width=1.0\textwidth]{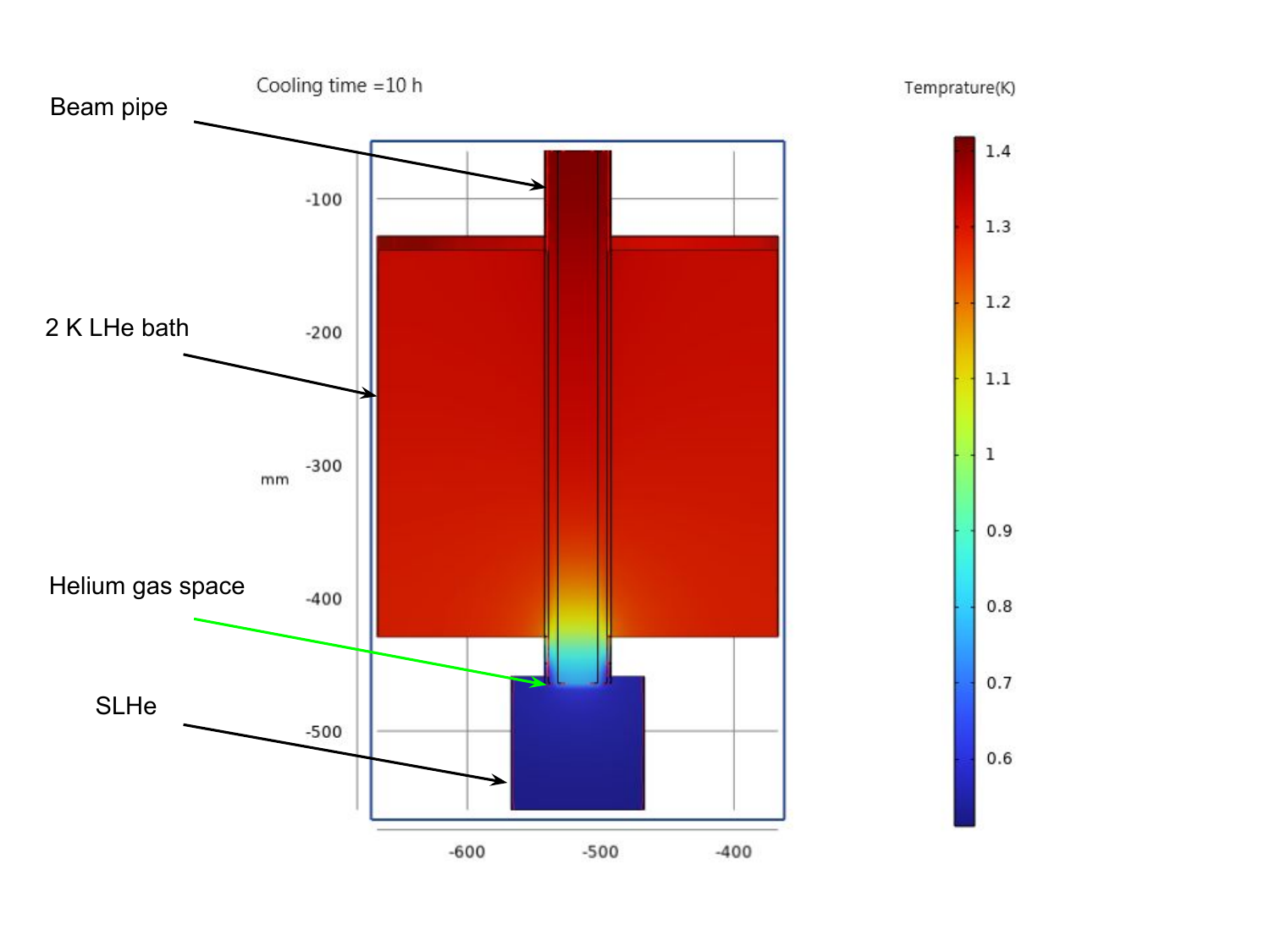}
	\caption{A COMSOL simulation showing the temperature of the helium gas space remains to be $\sim$ 0.5 K after 10 hours of cooling.}\label{COMSOL-Sim} 
\end{figure}

Our COMSOL~\cite{COMSOL} simulation aims to know the temperature of the 0.1 Pa helium gas space. In the simulation, we assume the SLHe's temperature is fixed at 0.5 K, and the LHe bath is 2 K. The simulation shows that after 10 hours of cooling, the 1 cm helium gas space is $\sim$ 0.5 K, Fig~\ref{COMSOL-Sim}, justifying the area's gas pressure is 0.1 Pa, as we hypothesized in our Geant4 simulation.

\section{Calibrating other liquid noble gases detector with COMIMAC}\label{sec4}

The COMIMAC facility can also generate the ion beams of other noble gases such as argon, neon, and xenon~\cite{discussionWithCOMIMACGuys}. In principle, the facility can calibrate other liquid gas detectors. However, as mentioned in section~\ref{sec3sub2}, it is critical to have the 1 cm thick vacuum space to let the beam pass through; without the vacuum area, the beam will be absorbed. It is challenging for argon~\cite{Chen75, ArgonPhaseChangeNist} and xenon~\cite{Clark51, XenonPhaseChangeNist} to reach a vacuum pressure at their triple point temperatures. However, neon can achieve a vacuum pressure if the temperature closes to 16 K~\cite{Stull47, NeonPhaseChangeNist}. According to the ideal gas law, decreasing the temperature will decrease the gas pressure. So, if argon and xenon have been cooled enough to a specific temperature, the pressure would be close to a vacuum; therefore, the calibrating method should be applicable. 

\section{Discussions}\label{sec5}

Calibrating a detector's NR response with the COMIMAC facility would significantly decrease the uncertainty thanks to the precise energy of the beam ($\lesssim$ 4\% uncertainty), as demonstrated by the LPSC team~\cite{newsG2022}. However, the beam must travel in a vacuum before hitting the target detector. Helium and neon gas will become a vacuum at 0.5 K and 16 K, respectively. It is difficult for argon and xenon to reach a vacuum to implement the method.

One should notice that whether the beam is helium or neon, it can only penetrate the target detector a couple of micrometers. As a result, it can not calibrate a TPC deep enough like the D-D generator setup at LUX~\cite{Huang22} and LZ~\cite{LZ2023}. 
Nevertheless, the table-size COMIMAC equipment fits well with the R\&D tests on cm-size detectors in a lab; it could also serve as a complementary online calibrating method for an LHe or liquid neon TPC under certain circumstances, thanks to its precise energy. Finally, the beam can be tuned for a $\sim$ 100 Hz events rate, so the chance of events pile-up would be minimal. In summary, the calibrating method introduced in the paper could be complementary to the NR calibration of LHe detectors.


\bmhead{Acknowledgments}

We thank Prof. Daniel Santos and his team members at LPSC, Dr. Jean-Francois Muraz, and Dr. Olivier Guillaudin, for their insightful discussions and comments from 2019 to 2023. We also thank Prof. Shaoliang Wang at AnHui University in China for helpful discussions. In addition, we thank the anonymous reviewer, who helped us to improve the quality of the paper in a couple of aspects. Junhui Liao thanks the ``Yuanzhang'' funding of CIAE to launch the ALETHEIA program. This work has also been supported by National Natural Science Foundation of China (NSFC) under the contract of 12ED232612001001 and the ``Continuous-Support Basic Scientific Research Project''.

\bmhead{Data Availability Statement}
The datasets generated during and/or analysed during the current study are not publicly available due to a preliminary research stage but are available from the corresponding author on reasonable request.

\bibliography{LHeNRCalibrationCOMIMAC}


\end{document}